\documentstyle[11pt,aaspp4]{article}

\begin{document}

\title{Parsec-scale radio structures in the nuclei of four Seyfert galaxies} 

\author{Marek J. Kukula}
\affil{Space Telescope Science Institute, 3700 San Martin Drive,
Baltimore, MD 21218, U.S.A. and Institute for Astronomy, University of
Edinburgh, Royal Observatory, Blackford Hill, Edinburgh EH9 3HJ, U.K.,
kukula@stsci.edu}

\author{Tapasi Ghosh} 
\affil{Arecibo Observatory, P.O. Box 995, Arecibo, Puerto Rico, PR
00613, U.S.A., tghosh@naic.edu}

\author{Alan Pedlar}

\affil{NRAL, University of Manchester, Jodrell Bank, Macclesfield SK11
9DL, U.K., ap@jb.man.ac.uk}

\and

\author{Richard T. Schilizzi}
\affil{Joint Institute for VLBI in Europe, P.O. Box 2, 7990 AA,
Dwingeloo, the Netherlands and Leiden Observatory, P.O. Box 9513, 2300
RA Leiden, the Netherlands, rts@nfra.nl}

\begin{abstract}

We present 18-cm radio maps of four Seyfert nuclei, Mrk~1, Mrk~3,
Mrk~231 and Mrk~463E, made with the European VLBI Network
(EVN). Linear radio structures are present in three out of four
sources on scales of $\sim$100~pc to $\sim$1~kpc, and the 20-mas beam
of the EVN enables us to resolve details within the radio structures
on scales of $<10$~pc.  Mrk~3 was also imaged using MERLIN and the
data combined with the EVN data to improve the sensitivity to extended
emission. We find an unresolved flat-spectrum core in Mrk~3, which we
identify with the hidden Seyfert~1 nucleus in this object, and we also
see marked differences between the two highly-collimated radio jets
emanating from the core. The western jet terminates in a bright
hotspot and resembles an FRII radio structure, whilst the eastern jet
has more in common with an FRI source. In Mrk~463E, we use the radio
and optical structure of the source to argue that the true nucleus
lies approximately 1~arcsec south of the position of the radio and
optical brightness peaks, which probably represent a hotspot at the
working surface of a radio jet. The EVN data also provide new evidence for
a 100-pc radio jet powering the radio source in the Type~1 nucleus of
Mrk~231. However, the Seyfert~2 galaxy Mrk~1 shows no evidence for
radio jets down to the limits of resolution ($\sim 10$~pc). We discuss
the range of radio source size and morphology which can occur in the
nuclei of Seyfert galaxies and the implications for Seyfert
unification schemes and for radio surveys of large samples of objects.

\end{abstract}

\keywords{galaxies: active --- galaxies: individual (Mrk~1, Mrk~3, Mrk~231, Mrk~463) --- galaxies: jets --- galaxies: Seyfert --- radio continuum: galaxies}

\section{Introduction} 

\placetable{tbl-1}

Seyfert galaxies are the nearest and most numerous class of active
galaxy and, although they are radio-weak, the nuclei of many Seyferts
contain compact non-thermal radio sources which often account for a
large fraction of the total radio luminosity of the galaxy. Aperture
synthesis with instruments such as the VLA and MERLIN has resolved
these radio sources, in many cases revealing linear, highly-collimated
structures ({\it eg} Ulvestad \& Wilson 1984a,b, Pedlar et al. 1993)
and there is now little doubt that at least some Seyfert galaxies are
capable of producing radio jets which are qualitatively similar to
those found in radio-loud quasars and radio galaxies, albeit several
orders of magnitude smaller and less powerful.

Because of their proximity, Seyfert nuclei present us with the
opportunity to study jet structure on the smallest possible spatial
scales, with a level of detail which is impossible for more luminous,
but also more distant, quasars and radio galaxies. In this paper we
report on very long baseline interferometer (VLBI) observations of
four Markarian Seyfert galaxies, Mrk~1, Mrk~3, Mrk~231 and Mrk~463E
(see Table~1). Preliminary results of this project were presented by
Ghosh et al. (1994).

We assume $H_{0} = 75$~kms$^{-1}$Mpc$^{-1}$ throughout this paper. The
spectral index, $\alpha$, is defined according to the relation
$S\propto \nu^{\alpha}$, where $S$ is the flux density at frequency
$\nu$.

\subsection{Radio emission and the standard model of Seyfert nuclei}

Several studies have attempted to use radio observations of large
samples of objects to test models of Seyfert activity. In particular,
the radio properties are regarded as ideal for testing schemes which
attempt to unify Seyfert 1s and 2s via viewing angle
effects. According to this scenario all Seyfert nuclei are surrounded
by an optically thick torus which lies in the plane perpendicular to
the axis of the radio jets. When our line of sight to the nucleus
intersects the torus, the central engine and broad line region (BLR)
are blocked from view and we classify the object as a Seyfert 2. If
our view of the nucleus is unobstructed, we call the object a Seyfert
1. Assuming that the radio emission is isotropic and unaffected by
dust obscuration, the model predicts that the average radio
luminosities of Types 1 and 2 should be the same and also that the
radio jets in Seyfert 1s should be foreshortened on average relative
to those in Seyfert 2s due to projection effects.

To date, radio surveys of Seyferts have produced conflicting results,
some of which can be ascribed to selection effects in the samples
themselves ({\it eg} due to the greater difficulty in identifying
Seyfert 2s; on average those that are detected tend to be more
luminous than Seyfert 1s). However, even when such effects have been
removed, or compensated for as far as is possible, confusion still
remains. Kukula et al. (1995) found that the radio structures of
Seyfert 1s in the CfA sample (Huchra \& Burg 1992) were on average too
compact to be be foreshortened versions of the Type 2 objects in the
same sample, indicating that orientation is not the only effect at
work in this particular case. Indeed, only 1/26 Seyfert~1s in the CfA
sample showed any sign of extended radio structure down to the
0.24$''$ resolution of the VLA.  Meanwhile, Roy et al. (1994), in a
single-baseline VLBI study of several Seyfert samples, found that
Type~1 objects were {\it less} likely than Type~2s to contain compact
nuclear radio components, contrary to what would be expected from the
simple unified model. Finally, Thean et al. (1998) report that Seyfert
1s and 2s in the Extended 12$\mu$m AGN Sample (Rush, Malkan \&
Spinoglio 1993) are {\it equally likely} to contain extended radio
sources, with no significant difference between the size distributions
of the two types. 

In practice the effects of the various selection criteria and
observational biases operating in these samples are very difficult to
quantify without a better understanding of the physical processes
which are responsible for the radio emission.  Detailed studies of
individual Seyferts offer perhaps the best means of achieving this
goal, and VLBI - though an inefficient tool for large surveys - is
ideally suited to this kind of work.

\subsection{VLBI observations of Seyfert nuclei}

VLBI provides the highest angular resolution of any imaging technique
but in the past sensitivity constraints were extremely tight,
effectively limiting VLBI studies of active galaxies to the small
fraction of objects which are radio loud. However, technological
advances made over the last decade have resulted in large increases in
sensitivity, and imaging of the brightest radio-quiet objects is now
well within the capability of modern instruments.

Several recent VLBI studies of Seyfert nuclei have provided detailed
images of structures close to the central engine itself (eg Halkides,
Ulvestad \& Roy 1997; Mundell et al. 1997; Oosterloo et al. 1998;
Ulvestad, Wrobel \& Carilli 1998; Roy et al. 1998).  Often the radio
structure on milliarcsecond (parsec) scales is misaligned with the
large-scale ($\sim 100$~pc) radio jets familiar from low-resolution
maps. These discoveries pose challenges for traditional unified models
of Seyferts because they imply that the aligned arcsecond-scale radio
continuum and optical line emission cannot necessarily be used to
trace the symmetry axis of the central engine.

However, in some cases the physical properties of the VLBI radio
source suggest an entirely different origin to the emission on larger
scales. In the nucleus of NGC~1068, the archetypal Seyfert~2 galaxy,
Gallimore, Baum \& O'Dea (1997) found a linear radio structure $\sim
1$~pc across, lying almost perpendicular to the arscecond-scale
synchrotron jet. They suggest that this feature represents free-free
emission from the inner edge of a molecular torus surrounding the AGN
- the first direct observational evidence for such a structure.
Ulvestad, Roy, Colbert \& Wilson (1998), consider a similar
explanation for the misaligned sub-parsec radio structure in NGC~4151.

\section{Observations and data reduction}

Observations were made in Autumn 1990 (see Table~1 for dates) with the
European VLBI Network (EVN), a radio interferometer spanning Western
Europe and at this time incorporating telescopes in Germany
(Effelsberg), Sweden (Onsala), Italy (Medicina) and the U.K. (Jodrell
Bank). The observing frequency was 1655~MHz (18~cm), giving an angular
resolution in the final maps of $\sim20$~mas - a factor of 2.5 higher
than that of the restored Hubble Space Telescope. The data recording
was in MK{\sc iii} mode~B with a bandwidth of 28~MHz. Two calibration
sources, OQ208 and DA193, were interleaved with the four target
sources during the 4 days of the observing run, with approximately 10
hours spent on each target source.  The amplitude calibration was
carried out using the system temperatures measured before and after
each 13-minute scan and the gain curves supplied by each
station. Global fringe-fitting was performed on the calibration
sources using point source models, and the resulting single-band and
multi-band delays were interpolated to the entire data-set.  The data
for each source were then fringe-fitted again and delays and rates
were adjusted before frequency and time averaging were performed.  For
Mrk~3, the previously made 18-cm MERLIN map (see below) was used for
the model, both for global fringe fitting and for the first round of
self calibration. For the other three sources, point source models
were used and hence the absolute position information has not been
retained. The calibration sources were also processed through several
cycles of phase-self-calibration and mapping. The resulting flux
densities for the two sources at this epoch of observation were
$S_{OQ208} = 0.8$~Jy and $S_{DA193} = 1.8$~Jy. We estimate an
uncertainty of $\sim 10\%$ in the flux densities.  For the final maps,
a uniform weighting scheme was used for the Fourier transform in order
to maximize the angular resolution. The RMS noise levels were
typically 0.15 - 0.2~mJy.

Eight hours of additional 18-cm (1658~MHz) observations were made of
one object, Mrk~3, on April 26$^{\rm th}$ 1993 with the Multi-Element
Radio Linked Interferometer Network (MERLIN), operated from Jodrell
Bank.  The MERLIN and EVN data were calibrated separately and the
visibilities then combined, with appropriate weighting, to give a
single large dataset. The MERLIN data helps to fill the `hole' in the
center of the $uv$ plane caused by the EVN's lack of short
baselines. After the Fourier transform and deconvolution, the
resulting map retained both the high angular resolution of the EVN and
MERLIN's greater sensitivity to extended radio emission.

\section{Results} 

Compact radio components were detected in all four Seyferts, with
three objects displaying prominent linear radio structures.  The
measured positions, flux densities, sizes and various derived
properties of the individual radio components are given in Table~2
(Mrk 1, 231 \& 463) and Table~3 (Mrk~3). Flux densities and positions
were measured from the maps using the {\sc aips} task {\sc
tvstat}. Deconvolved angular sizes for the components were obtained
from two-dimensional Gaussian fits to the data using the task {\sc
jmfit} or, for the more irregular features, were estimated from the
maps.

Estimates of the magnetic field density in the radio-emitting regions
derived from minimum energy arguments ({\it eg} Miley 1980), give
typical values of $\sim 10^{-7}$T (Tables~2 \& 3). The corresponding
lifetime for synchrotron-emitting electrons in a field of this
strength is $\sim10^{4}$~years. Brightness temperatures, $T_{B}$, of
the compact radio features are typically $10^{6-7}$K, similar to the
values measured in other Seyfert nuclei with the VLBA (Gallimore et
al. 1997: Ulvestad et al. 1998; Carilli et al. 1998).  However, the
spatial resolution of the current maps is insufficient to determine
whether parsec- or subparsec-scale free-free emission from a molecular
torus might be present in any of the objects in our sample.

\placetable{tbl-2}
\placetable{tbl-3}

\subsection{Markarian~1} 

\placefigure{f1}
 
This Seyfert 2 galaxy is perhaps best known for the prominent water
maser in its nucleus (Braatz, Wilson \& Henkel 1994).  Keel (1996)
suggests that the galaxy is interacting with the nearby object
NGC~451. No evidence for broad lines - indicative of a `hidden'
Seyfert~1 nucleus - has been found in Mrk~1, either in polarized light
(Kay 1994) or in the infrared (Veilleux, Goodrich \& Hill 1997).

Mulchaey \& Wilson (1995) report that the [O{\sc iii}]$\lambda5007$
emission in Mrk~1 is extended along PA83$^{\circ}$, whilst
Ulvestad, Wilson \& Sramek (1981), using the VLA at 6~cm, found the
radio structure to consist of a barely resolved nuclear source with
weaker emission extending $\sim400$~mas ($\sim 120$~pc) to the
south.

In the EVN map (Figure~1) we find a compact radio core (deconvolved
size $\sim10$~mas~$= 3$~pc) surrounded by a halo of emission
approximately 100~mas ($\sim 30$~pc) across. This extended emission
appears to be elongated in approximately the same direction as the
optical emission line region, although we note that the restoring beam
also has a similar orientation.  There is also some evidence for weak
emission extending to the south, perhaps leading into the larger
structure reported by Ulvestad, Wilson \& Sramek. We find no evidence
for linear radio structure in this object on any scale larger than a
few parsecs.

\subsection{Markarian~3} 

\placefigure{f2}

Mrk~3 is an early-type (S0 or elliptical) galaxy with a Seyfert 2
nucleus, although the presence of broad emission lines in its
polarized spectrum argues strongly for a hidden Seyfert~1 nucleus in
this object (Schmidt \& Miller 1985; Miller \& Goodrich 1990, Tran
1995), as does the photon budget calculated by Capetti et al. (1995a).
It is the nearest object in the current study; at a distance of only
55~Mpc we obtain a spatial resolution of 5~pc with the EVN.  In an
earlier study of this object with MERLIN at 6~cm, we found a
highly-collimated `S'-shaped radio structure with a bright radio
`hotspot' at its western end (Kukula et al.  1993). HST imaging of the
Narrow Line Region shows that the jet is surrounded by a sheath of
line-emitting gas (Capetti et al. 1995a), but there is no direct
one-to-one correspondence between the individual radio components and
brightness peaks in the line emission.

By concatenating 18-cm data from the EVN and MERLIN, with suitable
weighting of the data, we retain both the high angular resolution of
the former and the sensitivity to more extended emission provided by
the shorter baselines of the latter. The result, shown in Figure~2,
reveals a wealth of detail, with both the jets and the western
hotspot containing complicated substructure. In Figure~2, we adopt the
same naming scheme for the radio components as used to describe our 6-cm
MERLIN map, with the addition of component 1a, at the eastern end of
the structure (which was only marginally detected in the MERLIN map),
and with the refinement of letters to denote the most prominent
small-scale features.

The jets consist of a string of compact knots encased in a streamer of
more diffuse emission with a marked `S'-shaped curvature. There
appears to be no systematic increase or decrease in the brightness of
the knots along the length of the jet. The jet is resolved in the
transverse direction and its width is fairly constant ($\sim 100$~mas)
along its length. However there are several regions in which there
appears to be little or no radio emission: between the weak, diffuse
components 1a \& 1; on either side of the unresolved component 4; and
between the jet and the bright western `hotspot'. 

\subsubsection{Spectral index variations along the jet}

\placefigure{f3}

In order to study the variation in spectral index across the radio
source we convolved our 18-cm EVN map with a $0.09''\times0.07''$ beam
to make it directly comparable to our previous 6-cm MERLIN map (Kukula et
al. 1993). Using the {\sc aips} task {\sc comb} to compare the two
datasets resulted in the spectral index map reproduced in
Figure~3. Spectral indices for the individual components are listed in
Table~3.

The map shows the jet to have a steep spectrum, with a typical
spectral index of $\alpha \sim -1$. The compact features generally
have a slightly flatter spectrum than the diffuse emission, 
although the only knot with a genuinely flat spectrum is the unresolved
component~4.

The proximity of component~4 to the peak of the optical continuum
emission (as measured by Clements 1981), and its position at the
center of symmetry of the `S', first led us to suggest that it might
be the radio core of Mrk~3 (Kukula et al. 1993). The newly-derived
flat spectrum for this component supports this view since it is
characteristic of synchrotron self-absorption (SSA), a property which
is often observed in the core components of other AGN.  If this is the
case then the jet in Mrk~3 is two-sided and we see no emission on
either side of the nucleus for the first $\sim 30$~pc.

If SSA is indeed responsible for the flat radio spectrum of this
component then its brightness temperature must be in excess of
$10^{9}$K, although at present we can only place a lower limit of
$3.5\times10^{6}$K on $T_{B}$ (Table~3). In order for the core to be
synchrotron self-absorbed it must have a maximum size of $\sim 2$~mas
($\sim 0.5$~pc). This prediction will be tested by forthcoming VLBA
observations.

\subsubsection{Interpretation of the radio structures in Mrk~3}

The radio source in Mrk~3 is one of the nearest known examples of an
AGN containing a pair of highly-collimated radio jets. The details
revealed in our new map of Mrk~3 therefore show the structure of such
systems on a very small spatial scale. Here we discuss the nature of
the jets in Mrk~3 based on this small-scale evidence.

{\bf General structure:} As noted previously, the radio emission
follows an overall `S'-shaped distribution. This could be due either
to a change in the ejection axis of the jets with time, or else to the
interaction of the jets with a rotating interstellar medium. As
Figure~4 shows, the brightest optical line emission tends to be
associated with the leading (convex) edge of the `S' (Capetti et
al. 1995a) - an observation which could favor either scenario.


{\bf The flat-spectrum core:} In radio-loud sources the innermost
radio components observed with VLBI are usually assumed to mark the
first recollimation shocks within the jet. The identification of an
unresolved flat-spectrum component in Mrk~3 therefore provides us with
the best estimate of the position of the central engine. The
deconvolved size of this source is $<10$~pc, suggesting that the first
shocks in the jet must occur within 5~pc of the central
engine. Clearly an immediate goal for future VLBI would be to study
this region in more detail and hopefully to resolve the core component
into two recollimation shocks. Flux density comparisons might then
give some clues as to the presence of relativistic velocities in the
inner jet regions and the orientation of the two jets with respect to
our line of sight.

{\bf The western hotspot:} The hotspot itself appears to be
edge-brightened, but with a complex of bright knots in the center. The
overall appearance is suggestive of a shell or bowshock where the jet
is interacting violently with the external medium. The knots could be
the result of radio emission over a relatively large 3D interaction
surface at the head of the jet which is seen in
projection. Alternatively they could mark a series of increasingly
violent collisions as the jet is diverted within the hotspot before
finally impacting on the external medium.

\placefigure{f4}

{\bf A comparison of the western and eastern jets:} Apart from their
participation in the overall `S'-shaped morphology of the source, the
two radio jets display a number of obvious differences. The
most striking of these is the absence at the end of the eastern jet
of strong radio emission and of any structure analogous to the bright
western hotspot. Another obvious difference is that, unlike its western
counterpart, the eastern jet contains two bright radio components
within 100~pc of the nucleus.

It is tempting to suggest that these two circumstances are related.
If, as suggested above, the radio knots mark the sites of shocks
where oscillations in the jet direction have caused it to strike the
walls of the channel, then the presence of two strong collisions early
on in the course of the eastern jet might cause much of the bulk
kinetic energy of the jet material to be thermalised, leading to a
large drop in the jet's Mach number.

Radio maps at lower resolution (eg Unger et al. 1986; Kukula et
al. 1993) show that beyond the structure visible in the current maps
the eastern jet emission curves to the south and continues to fade.
Interestingly, the HST study by Capetti et al. (1995a) shows that the
line emission in this region {\it does not} fade but remains bright
(Figure~4). However, the ionisation state here, as traced by the
[O{\sc iii}]/H$\alpha$ ratio, is two to three times higher than in the
rest of the NLR, indicating lower gas densities. By contrast, the
diffuse line emission associated with the western hotspot is
relatively faint and has a low ionisation, which Capetti et
al. interpret as evidence for a high-density shell of shocked material
surrounding the radio lobe.

The differences in the properties of the two jets recall those between
the two main types of structure found in classical radio galaxies:
FRII sources, in which supersonic jets end in sharp-edged hotspots
confined by shocks; and FRIs, where the jets appear to become subsonic
before they terminate (see, for instance, Leahy 1991).  Thus the
western jet in Mrk~3 is analogous to an FRII jet, maintaining a high
Mach number along its length and terminating in a violent shock where
it impinges on the external medium.  In contrast, the eastern jet,
once it has passed through the two bright knots, behaves more like an
FRI source, dissipating into its surroundings without producing strong
shocks.  Thus, the radio source in Mrk~3, although roughly an order of
magnitude less luminous than classical radio sources and confined to
the nuclear regions of its host, might offer insights into both types
of structure displayed by large, powerful radio galaxies. Further
high-resolution work on this object, such as our own forthcoming VLBA
observations, will be necessary in order to determine the spectral
indices of the compact radio features more accurately, and to search
for proper motions and other changes within the jet.
 
\subsection{Markarian~231} 

\placefigure{f5}

Mrk~231 is an ultra-luminous infrared galaxy (Sanders et al. 1988), 
whose Seyfert 1 spectrum also contains multiple blueshifted
absorption-line systems (Boksenberg et al. 1977).  Boksenberg et
al. also estimate 2 magnitudes of extinction towards the optical
nucleus of Mrk~231, implying that the object is technically a
radio-quiet quasar rather than a Seyfert galaxy, according to the
luminosity criterion of Schmidt \& Green (1981).

Mrk~231 possesses complex radio structure on a variety of physical
scales, consisting of both compact, high-surface-brightness features
and regions of more diffuse emission. At the lowest angular
resolutions radio images of Mrk~231 show a bright, nuclear source with
a large region of faint emission extending $\sim 30$~arcsec ($\sim
2.4$~kpc) to the south (Hutchings \& Neff 1987; Carilli, Wrobel \&
Ulvestad 1998). The nuclear radio source remains point-like at VLA
resolutions ({\it eg} Kukula et al. 1995) but with VLBI the component
is resolved into a small, high-surface-brightness central
component embedded in a diffuse, roughly elliptical region of emission
$\sim 200$~mas ($\sim 160$~pc) across (Carilli et al. 1998). Carilli
et al. interpret this diffuse structure as synchrotron emission from
the inner part of the molecular gas disc detected in CO by Bryant \&
Scoville (1996).

The compact radio source at the center of the `disc' has a convex
radio spectrum, and a high brightness temperature (McCutcheon \&
Gregory 1978; Ulvestad, Wilson \& Sramek 1981; Neff \& Ulvestad 1988).
Neff \& Ulvestad (1988), using three-station VLBI with the EVN in the
early 1980s, were able to demonstrate that this component is elongated
in a north-south direction, perpendicular to the major axis of the
molecular disc. More recent VLBA observations have further resolved
the source into a core-lobe structure (Ulvestad, Wrobel \& Carilli
1998; Carilli et al. 1998) with a bright central component and
southern lobe separated by $\sim 30$~mas ($\sim 20$~pc), and a weaker
lobe 20~mas to the north. At high frequencies the `core' itself is
resolved into a 2-milliarcsec (1.5-pc) triple source which is
misaligned with the `lobes' to the north and south (Ulvestad, Wrobel
\& Carilli 1998), reminiscent of the parsec-scale radio structures of
NGC~1068 (Gallimore et al. 1997) and NGC~4151 (Ulvestad et al. 1998).

The lack of short baselines in the EVN means that it is most sensitive
to radio structures which are intermediate in size between the
small-scale ($\sim 30$~mas) core-lobe source and the larger {$\sim
200$~mas) `disc' detected with the VLBA by Carilli et al. (1998). In
the current EVN map of Mrk~231 (Figure~5) the diffuse `disc' emission
is resolved out, leaving only the linear nuclear source. The core-lobe
structure from the VLBA maps is only partially resolved by the EVN and
forms the brightest part of a somewhat larger ($\sim 100$~pc) jet-like
structure aligned in PA$\sim20^{\circ}$. In addition to the VLBA
source, this `jet' contains two more radio components $\sim 70$~mas
(50~pc) and 130~mas (90~pc) to the southwest.

The total flux density of the VLBA core-lobe source does not appear to
have varied significantly during the six years separating the EVN and
VLBA observations ($94\pm 10$~mJy at 1.6~GHz with the EVN in September
1990, and $100\pm 10$~mJy at 1.4~GHz with the VLBA in December
1996). Our measured fluxes are also in good agreement with those of
Lonsdale, Smith \& Lonsdale (1993), made in September 1991 at
1.5~GHz. At higher frequencies, where the emission from the extended
steep-spectrum components is less dominant, the radio source is known
to be variable (McCutcheon \& Gregory 1978; Ulvestad, Wilson \& Sramek
1981; Neff \& Ulvestad 1988).  


Clearly, our understanding of this object is far from complete.  The
presence of structures on a variety of scales and in different
orientations - the $\sim160$-pc `disc' identified by Carilli et
al. (1998), the $\sim50$-pc `jet' seen in the current study, and the
misaligned $1.5$-pc nuclear triple source detected with the VLBA by
Ulvestad, Wrobel \& Carilli (1998) - suggest that several processes
are contributing to the radio continuum emission in Mrk~231.
Ulvestad, Wrobel \& Carilli (1999) provide a more detailed discussion
of the various radio components in this object and the relationships
between them.


\subsection{Markarian~463E}  

\placefigure{f6}

\placefigure{f7}

This object forms the eastern component of an interacting pair of
galaxies.  Its Seyfert 2 spectrum contains polarized broad H$\alpha$
(Miller \& Goodrich 1990, Tran 1995) and, in the near infrared, broad
Pa$\beta$ (Veilleux, Goodrich \& Hill 1997), indicating the presence of
a hidden Seyfert 1 nucleus. Mazzarella et al. (1991) infer large
amounts of visual extinction towards the nucleus and speculate that,
but for this, Mrk~463E would be classified as a quasar.  Ground-based
studies of the [O{\sc iii}] emission show that the nuclear region is
elongated in PA$\sim180^{\circ}$ and that fainter emission fills a
roughly biconical volume extending $\sim20$~arcsec ($\sim$18~kpc) to
the north and south of the nucleus (Hutchings \& Neff 1989).

The object possesses radio structure on several different scales, all -
like the [O{\sc iii}] emission - aligned in a roughly north-south
direction. Mazzarella et al. (1991) report radio components 4~arcsec
(3.6~kpc) north and 18~arcsec (16.2~kpc) south of the nucleus. Strong
radio emission is also associated with the nucleus itself: existing
radio images (Unger et al. 1986, Neff \& Ulvestad 1988, Mazzarella et
al. 1991) show a bright, slightly extended radio source coincident
with the optical brightness peak of the galaxy, with a second radio
component lying $1.2$~arcsec (1~kpc) to the south. In terms of both
luminosity and overall size the radio source associated with Mrk~463E
is intermediate between the sources typical of the majority of Seyfert
galaxies and those found in radio galaxies and quasars.

Optical continuum imaging with the F517N and F547N filters of the HST
(Uomoto et al. 1993) resolves the nuclear region, revealing a bright
component $\sim 500$~mas across, with an optical `jet' extending
southwards for 820~mas (740~pc) and stopping just short of the
southern radio component. Uomoto et al. attribute this radiation to a
mixture of emission lines produced {\it in situ} and scattered light
(both continuum and emission-line) from the hidden Seyfert~1 nucleus,
rather than optical synchrotron emission.

At a distance of $\sim 200$~Mpc, the 20-mas EVN beam corresponds to a
physical size of 18~pc.  In our new map (Figure~6) the brightest radio
source is resolved into four discrete components, all aligned in
roughly the same direction as the large-scale structure. Three of the
sources form a closely spaced triplet, whilst the fourth lies some
300~mas (270~pc) further south.  The radio emission to the south of
the optical `jet' is also resolved, and is elongated in the same
direction as the overall radio and optical structures. The radio
components become progressively weaker as one moves southwards.

\subsubsection{A new interpretation of the nucleus of Mrk~463E}

The optical brightness peak of Mrk~463E has a Type~2 (narrow line)
spectrum and has traditionally been assumed to mark the site of the
supermassive black hole responsible for the activity in this
object. Since the brightest radio component in Mrk~463E is roughly
coincident with the optical peak it followed that this was the radio
source associated with the obscured central engine. However, in other
Seyferts with prominent linear radio structures this is not always the
case; in both Mrk~3 (Kukula et al. 1993; this paper) and Mrk~6 (Kukula
et al. 1996) the brightest radio component occurs at the end of the
jet furthest from the nucleus, is resolved, edge-brightened, and has a
steep radio spectrum - all the hallmarks of a `hotspot' where the jet
material is ploughing into the ISM of the host galaxy. In both cases,
the radio emission is associated with enhanced narrow-line
emission, which appears to surround the radio structure like a shell
or halo.  In Mrk~3 the hidden central engine is associated with an
unresolved, flat-spectrum radio component. In Mrk~6, the Seyfert~1
nucleus does not appear to coincide with any of the detected radio
components.  Roy et al. (1994) invoked free-free absorption by ionized
gas in the narrow line region to explain the lack of compact nuclear
radio sources in their sample of Seyfert~1s (in Seyfert~2s we observe
the object `side on' and so the depth of ionized gas between the
observer and the central engine is much reduced). The same mechanism
might also account for the absence of a radio component at the
position of the Type~1 nucleus in Mrk~6.

Tremonti et al. (1996) used HST imaging polarimetry to locate the
hidden Seyfert~1 nucleus in Mrk~463E. They report that the magnetic
polarization vectors converge on the southern tip of the optical
`jet', indicating that this, and not the bright narrow-line region
north of the `jet', is the true site of the central engine in this
object.

The detailed radio structure of Mrk~463E revealed by our new EVN
observations, when considered in conjunction with the optical
structures reported by Uomoto et al. (1993), lends strong support to
this finding. In Figure~7 we show the current 18-cm radio map
(greyscale) superimposed on the HST continuum image by Uomoto et
al. (contours). There is some question as to the positional accuracy
of the two images. The radio data has undergone self-calibration,
causing the absolute positional information to be lost. More
seriously, HST pointing is generally only accurate to within
$\sim1$~arcsec, and the true pixel scale of the HST Planetary Camera
image may be slightly different from the canonical value of
43.9~mas~pixel$^{-1}$. In performing the registration of the two
images we have forced the brightness peaks to coincide. Several
aspects of the argument set out below rely heavily on this assumption,
so it is clearly imperative that more accurate radio and,
particularly, optical astrometry be carried out on this object.

The overall outline of the optical emission in the HST image is
roughly wedge-shaped, with its apex at or near the southern radio
component, 1.2~arcsec from the presumed `Seyfert~2 nucleus'. The
enhanced optical emission forming the `jet' begins immediately north
of this radio component and continues north for almost a kiloparsec,
incorporating three bright knots, before terminating abruptly. A gap
of $\sim 0.1$~arcsec (90~pc) separates the `jet' from the bright,
resolved `nucleus' to the north and a second radio component appears
to lie in the gap.

Our suggested interpretation of Mrk~463E is that the `Seyfert~2
nucleus' at the northern end of the jet is in fact a region of
enhanced narrow-line emission associated with a radio hotspot. In
accordance with the data of Tremonti et al. (1996) we suggest that the
central engine itself lies at the southern end of the optical jet,
although for the reasons outlined below we feel that it is unlikely to
be exactly coincident with the southern radio component.  The overall
wedge-shape of the extended optical emission can be accounted for by
shadowing of the radiation field of the AGN by an optically-thick
torus surrounding the central engine - the favored model for several
other Seyferts with similar optical morphologies ({\it eg} Wilson \&
Tsvetanov 1994).

We interpret the linear optical feature seen in the HST image as an
expanding cylindrical sheath or halo of material surrounding the
channel of the radio jet. Such a model was suggested by Capetti and
co-workers to explain the very similar structures seen in emission
lines in several Seyferts, including Mrk~3 \& 6 (Capetti et al. 1995a,
\& b, 1996).  The material in the halo has been pushed aside by the
passage of the radio plasma, and is subsequently ionized by UV photons
from the central engine. The high density of the halo relative to the
surrounding gas also makes it a more effective site for scattering of
the nuclear continuum. Uomoto et al. (1993) claim that the F547M to
F517N flux ratio of the `jet' indicates a mixture of optical continuum
and emission lines. This can be explained in our model by
contributions from line emission produced {\it in situ} by the
swept-up, ionized gas, as well as polarized continuum and broad
permitted line emission (already detected by Miller \& Goodrich 1990)
from the obscured nucleus which have been scattered into our line of
sight by electrons in the halo. Double-peaked line profiles would be
indicative of an expanding cylindrical sheath of material. Free-free
continuum emission from shocked gas in the halo might also be
present. Optical spectroscopy and spectropolarimetry with high spatial
resolution will be required in order to determine the precise
contributions from each mechanism.

The estimated lifetime for $\lambda$~18cm synchrotron-emitting
electrons in this source (derived from the minimum energy magnetic
field values listed in Table~2) is $\sim10^{4}$~years. Hence, unless
the bulk velocity of the jet is an appreciable fraction of the speed
of light ($\geq 0.01c$) the electrons in the plasma will radiate most
of their energy before traveling more than a few tens of parsecs from
the origin of the jet. At larger distances a fresh injection of energy
will be required in order to produce appreciable amounts of
synchrotron emission.  We would therefore expect to detect strong
synchrotron emission at sites where the jet plasma is interacting
violently with its surroundings, leading to the re-acceleration of
electrons, but very little where the jet is flowing freely.

Thus, in the region of the optical `jet' the plasma has already
drilled out a channel for itself, through which it flows relatively
unimpeded, and produces little or no synchrotron emission.  The radio
emission immediately north of the linear optical feature might
indicate that the jet is colliding with an obstacle at this point. It is
not clear why the optical emission should simultaneously fall in
brightness, but Taylor, Dyson \& Axon (1992) point out that the shocks
produced by the collision of a jet with an external medium can heat
the swept-up material to temperatures $\gg 10^{4}$K, too large for the
emission of forbidden lines to occur.

Following this encounter, which appears to alter the course of the
jet, the plasma feeds into the bright radio hotspot. This is the site
at which the jet is actively carving a channel into the local ISM. The
observed radio structure of the hotspot consists of three successively
brighter components, the first two of which are compact and a third
which is resolved in a direction roughly orthogonal to that of the
jet. We suggest that these features are analogous to the internal
radio structure observed in the western hotspot of Mrk~3 (Figure~2)
and that the bright transverse feature is associated with a bowshock
at the head of the jet. Mazzarella et al. (1991) detected $\sim
320$~mJy of emission in this region with the VLA at 20~cm, implying
that over 100~mJy of extended flux has been `resolved out' by the
small beam of the EVN. The VLA map by Mazzarella et al. also contains
a weak radio component 4~arcsec (3.6~kpc) north of the nucleus. This
could indicate either that the jet is not entirely disrupted in the
structure which we have dubbed the hotspot, or else that the hotspot
was formed when dense material drifted into the path of the jet,
cutting off the radio structure further to the north.

Unger et al. (1986) found that the radio emission in Mrk~463E has a
steep spectrum between 408 and 1666~MHz. Although the angular
resolution of their maps was not sufficient to resolve the individual
knots which make up the northern component, a steep radio spectrum in
this region lends credence to our suggestion that this is a radio
hotspot rather than the central engine itself. The radio component to
the south of the optical `jet' also has a steep spectrum and the
emitting region is resolved in our EVN map. This makes it unlikely to
be directly associated with the central engine, since the nuclear
components observed in other Seyferts tend to be compact, with flat
spectra.

However, further evidence that the true nucleus is located in this
vicinity is the mention by Neff \& Ulvestad (1988) of a region of
extremely blue optical continuum emission apparently coincident with
the southern radio component (in our EVN map).  We suggest that this
blue emission might be a signature of the hidden Seyfert~1 nucleus.
New studies of the broad-line emission from Mrk~463E (polarized
H$\alpha$ in the optical and broad Pa$\beta$ in the near infrared)
with improved spatial resolution would also assist in defining the
scattering geometry and locating the central engine.

Mazzarella et al. (1991) also found a knot of radio emission 18~arcsec
(16.2~kpc) south of the nucleus in their 20~cm VLA image, which could
be construed as a radio hotspot at the end of a southern counterpart
of the jet in the north (although, of course, it might equally well be
entirely unrelated to the AGN).  The relative weakness of the radio
emission in this component, and the lack of any detectable emission
from the counterjet itself, might be explained in a number of ways.

If the counterjet is directed away from us, into the plane of the sky,
and the bulk velocity is relativistic then Doppler dimming could lead
to a reduction in the observed radio emission.  A difference in the
environment to the south of the nucleus might also lead to less radio
emission: if the counterjet encounters little resistance from the ISM
then there will be few opportunities for the re-acceleration of
electrons and a corresponding absence of synchrotron radiation. A
lower density in the southern ISM would also account for the greater
length of the counterjet (16~kpc) relative to the maximum extent of
the northern radio emission (3.6~kpc).  Certainly, in view of the
disturbed nature of the Mrk~463 system, we would not necessarily
expect the galaxy's density contours to be symmetrical about the
nucleus of Mrk~463E.

A final possibility is that the counterjet is currently dormant or
inactive and the southern radio component is therefore the fading remnant
of a more luminous hotspot.

The ground-based [O{\sc iii}] image by Hutchings \& Neff (1989) shows
a fan-shaped region of emission extending $\sim 20$~arcsec to the
north of the nucleus. However, to the south the [O{\sc iii}] emission
is concentrated in a bright, resolved knot at a distance of $\sim
10$~arcsec from the nucleus, with little extended emission. This could
indicate either larger amounts of obscuration between the Earth and
the ionized gas to the south (perhaps because the southern cone of
ionising photons is directed away from us) or else, once again, a lack
of gas in this region. 

However, at this stage we cannot rule out the possibility that much of
the faint [O{\sc iii}] and radio emission on arsecond scales is
associated with starforming regions rather than the central engine of
Mrk~463E.

To summarize: we suggest that the optical and radio brightness peak in
Mrk~463E has been mistakenly identified as the location of the active
nucleus in this object, when in fact it represents the working surface
of a jet, which originates $\sim 1$~arcsec to the south. Since our
understanding of this object depends crucially on the ability to match
up high-resolution images made in many different regions of the
spectrum, more accurate astrometry at all wavelengths is urgently
required.

\section{Discussion}

Three of the objects in our study (Mrk~3, 231 and 463E) were
observed previously with the EVN in the early 1980s at 1.6~GHz (Neff \&
Hutchings 1988). Our maps agree with the previous data in terms of
the overall morphology of the sources but advances in instrumentation
and software have resulted in significant improvements in image
quality and the new maps therefore contain much more detail.

Most of the compact radio emission in all four Seyferts has a steep radio
spectrum (component~4 in Mrk 3 has a flat spectrum) 
and is clearly non-thermal in origin. Neff \& Ulvestad (1988)
point out that unrealistically large supernova rates would be required
in order to reproduce the radio luminosities observed in the compact
radio components of Mrk~3, 231 and 463; based on our current
measurements the same appears to be the case for Mrk~1.  Moreover, the
combination of high ($\geq 10^{5}$K) brightness temperatures and steep
radio spectra appears to rule out supernovae as the primary source of
the synchrotron emission since, in a starburst complex, brightness
temperatures would be limited to $\leq 10^{5}$K and one would expect
the radio spectrum to flatten for $T_{B}>10^{4}$K (see Condon 1992).

An AGN is therefore the most likely source of the VLBI radio
emission, and in three of the objects we have mapped, the radio
morphology also strongly suggests the presence of less powerful forms
of the jets found in radio-loud sources. However, there is evidence
for galaxy interactions/mergers in all four cases, and the more
diffuse and amorphous radio emission (for example, that detected on
arcsecond scales in Mrk~231 by Hutchings \& Neff 1987, or the radio
halo in Mrk~1) may well be the result of starburst activity rather
than a direct product of the AGN.

What can be learned from these four objects? We cannot hope to draw
generalized conclusions about Seyferts as a class from such a small
sample. Also, optical imaging and kinematical studies indicate that
all four objects in our study are currently interacting with
neighboring galaxies (Mrk~1, 3 \& 463E) and/or have recently
undergone mergers (Mrk~3 \&~231). Though such processes may well play
a r\^{o}le in triggering and fueling the activity in all active
galaxies ({\it eg} Phinney 1994), the prominent disturbances in these
four galaxies might account for some of the complexity in their radio
structures.  However, the current observations do highlight a number
of important points which should be taken into account in any
consideration of Seyfert radio properties.

With very high angular resolution, jet-like structures {\it can} be
found even in compact Type~1 objects such as Mrk~231, in which the
axis of the central engine is probably aligned quite closely with the
line of sight.  The presence of large-scale ($>100$~pc) collimated
radio jets in Mrk~3 and 463E also serves to reinforce the message that
Seyfert~1s are capable of producing such structures since, although
these two objects have Type~2 spectra as observed from our own vantage
point, spectropolarimetry and infrared spectroscopy indicate that both
contain a hidden Type~1 nucleus.

On the other hand, there are Seyferts such as Mrk~1 which, although
they possess respectably luminous radio sources, show no signs of
jet-like structure on scales as small as a few parsecs. Mrk~1 is a
particularly good example of this class because, as a Type~2 Seyfert,
the unified model predicts that the ejection axis for a putative radio
jet should be close to the plane of the sky, thus minimizing any
projection effects. It is interesting to note that, in the present
sample, both of the Type~2 Seyferts which show evidence for a hidden
Type~1 nucleus also contain radio jets, whilst Mrk~1, in which all
attempts to find broad lines have so far failed, does not.

It appears that whilst some Seyfert galaxies of both types are
certainly capable of producing linear, jet-like radio sources, the
radio structures in Seyferts can range in size from extremely small
($<10$~pc) nuclear sources through to jets on scales of tens, hundreds
and even thousands of parsecs. This large distribution of sizes is
intrinsic to the Seyfert population and is quite separate from viewing
angle effects. Possibly the radio jets in Seyferts are relatively
short-lived compared to the lifetime of the active nucleus, so that
when we observe a sample of objects, we see jets in various stages of
development, as well as objects in which jets are not currently
active. Alternatively, there may be some Seyferts of both spectral
types which are simply incapable of producing radio jets, perhaps due
to environmental factors such as the gas density/kinematics in the
nuclear region of the host galaxy and the orientation of the central
engine relative to the plane of the host.

The factors at work in the host galaxy (dust and gas content,
interaction and star-formation history) and the active nucleus
(immediate environment and orientation of the central engine, the
size, age and morphology of the radio source), as well as the biases
and omissions introduced by observational constraints, combine to make
the observed properties of a particular Seyfert galaxy distinctive and
unique. This point is illustrated by the cases of Mrk~3 and 463E, in
which jet and counterjet have very different appearances due, perhaps,
to the different conditions which they encounter {\it en route}.  One
should not forget that Seyfert galaxies are complicated systems
composed of many elements, whose properties can change and evolve on
relatively short timescales.  Consequently one should not be too
surprised if simplistic models fail on some level when applied
globally to large samples of objects.

Further studies of the relationship between the AGN and its host need
to be carried out before we can be confident about our interpretation
of Seyfert radio surveys. High-resolution imaging at many wavelengths
will clearly play a major r\^{o}le in such studies, but, as the present
work demonstrates, accurate astrometry will be crucial in order to
match up the different datasets.

\section{Summary} 

The maps presented here clearly demonstrate that VLBI can be a highly
effective tool for investigating the radio structures of Seyfert
nuclei, even though the radio emission from these objects is
relatively weak when compared to radio-loud sources. Since Seyferts
are in general much closer to us than other types of AGN, the spatial
resolution achieved by VLBI is unprecedented, allowing us to study the
structure of the radio jets on scales of a few parsecs and to
investigate in detail the complex interactions between the jets and
their environment.

Radio jets exist in both broad- and narrow-line Seyferts, although the
scale of such structures varies from a few tens of parsecs to several
kiloparsecs. The range in sizes appears to be intrinsic to the Seyfert
population and is independent of any additional projection effects
caused by the orientation of the central engine. However, there are
also objects which possess luminous nuclear radio sources yet which
show no evidence for jets, even on scales of $\leq10$~pc.

In Mrk~3 and 463E we find evidence for jets containing internal
shocks and terminating in luminous, edge-brightened hotspots
resembling the structures seen in large-scale FRII radio
sources. However, in general the observed radio properties of an
individual Seyfert nucleus are likely to be a complex function of the
local environment as well as of the central engine itself, making
each object unique. This has important consequences for the
interpretation of radio surveys of Seyferts.

\acknowledgements 
The authors would like to thank Bill Junor for help in reducing the
data, Ant Holloway for producing Figure~4, Christina Tremonti and Alan
Uomoto for useful discussions on the subject of Mrk~463, and Ger de
Bruyn, George Miley and Dave Graham for their invaluable contribution
to the early stages of this project. We are also grateful to James
Ulvestad, the referee, whose comments resulted in significant
improvements to this paper. MJK acknowledges PPARC support and STScI
funding (grant number O0573). The EVN is a large-scale facility of the
European Union and is administered by the European Consortium for
VLBI.  MERLIN is a national facility operated by the University of
Manchester on behalf of PPARC.  This research has made use of the
NASA/IPAC Extragalactic Database (NED) which is operated by the Jet
Propulsion Laboratory, California Institute of Technology, under
contract with the National Aeronautics and Space Administration.

\clearpage


\clearpage
 
\figcaption[f1.eps]{18-cm EVN map of Mrk~1. The beam is $0.024 '' \times 0.018
''$ in PA $71^{\circ}$ (indicated by the ellipse in the lower left
hand corner of the map), equivalent to a spatial resolution of
$\sim6$~pc. (Peak flux density $=$ 31~mJy~beam$^{-1}$; $n^{th}$
contour is $0.14 \times 2^{n-1}$~mJy~beam$^{-1}$.)
\label{f1}}

\figcaption[f2.eps]{Top: 18-cm EVN+MERLIN map of Mrk~3. The beam is $0.027''
\times 0.023''$ in PA$51^{\circ}$, equivalent to a spatial resolution
of $\sim5$~pc. Peak flux density $=$ 34.4~mJy~beam$^{-1}$; 3$\sigma$
noise is 0.2~mJy~beam$^{-1}$, with contours at -1, 1, 5, 10, 15, 20,
25, 30, 40, 50, 75, 100, 150 \& 300 times this value. The cross marks
the position of the optical nucleus as determined by Clements
(1981). Bottom: Cartoon of the radio emission in Mrk~3, showing the
distribution of the extended emission (solid lines) and the most
prominent compact radio features (shaded regions). The numbering
scheme is modified from that used to describe the original 6-cm MERLIN
map (Kukula et al. 1993).
\label{f2}}

\figcaption[f3.ps]{Spectral index map of Mrk~3. The spectral index, $\alpha$
(defined by $S \propto \nu^{\alpha}$), is indicated by the greyscale
bar at the top of the frame.  The measurements were made between 1.6
and 4.8~GHz by comparing the current EVN$+$MERLIN data with the
4.8-GHz MERLIN map by Kukula et al. (1993). The two maps were
convolved with an identical $0.09''\times0.07''$ beam.  The cross
marks the peak of the optical emission, as determined by Clements
(1981).
\label{f3}}

\figcaption[f4.ps]{[O{\sc iii}] emission in the nucleus of Mrk~3
(greyscale) imaged with the HST by Capetti et al. (1995a) through the
F501N filter, with the 18-cm radio emission (contours)
superimposed. Note the bright knot of [O{\sc iii}] to the south of the
eastern radio jet, and also how the line emission appears to favour
the convex edge of the S-shaped curvature.  As before, the cross marks
the peak of the optical continuum emission, as determined by Clements
(1981), which appears to be offset from the hidden Seyfert 1 nucleus by
$\sim 200$~mas.
\label{f4}}

\figcaption[f5.eps]{18-cm EVN map of Mrk~231. At a distance of 164~Mpc the
$0.029''\times 0.021''$ (PA$54^{\circ}$) beam is equivalent to a
spatial resolution of $\sim15$~pc. (Peak flux density $=$ 75~mJy~beam$^{-1}$;
n$^{th}$ contour is $0.7 \times 2^{n-1}$~mJy~beam$^{-1}$.)
\label{f5}}

\figcaption[f6.eps]{18-cm EVN map of Mrk~463E. The beam has a FWHM of $0.023''
\times 0.022''$ in PA$49^{\circ}$, equivalent to a spatial resolution
of $\sim18$~pc. (Peak flux density $=$ 38~mJy~beam$^{-1}$; $n^{th}$
contour is $0.6 \times 2^{n-1}$~mJy~beam$^{-1}$.)
\label{f6}}

\figcaption[f7.ps]{18-cm EVN map of Mrk~463E (greyscale; scale bar marked in
mJy~beam$^{-1}$) with the HST optical continuum (F517M filter) image
by Uomoto et al. (1993) superimposed (contours). The registration of
the two datasets was performed by forcing the optical and radio
brightness peaks to coincide. The radio map has been convolved with a
gaussian function to give it a comparable resolution to the
HST image ($\sim 50$~mas).
\label{f7}}

\clearpage

\begin{deluxetable}{lcrcccl}
\tablecaption{Objects in the current sample. Optical positions are
taken from Clements (1981) and refer to the peak of the optical
continuum emission. Dates for the 18-cm EVN observations reported in
this paper are listed as day/month/year (additional 18-cm observations
of Mrk~3 were made with MERLIN on 26/04/93).
\label{tbl-1}}
\tablewidth{0pt}
\startdata
Object & Seyfert & \multicolumn{1}{c}{$cz$} & \multicolumn{2}{c}{Optical position of nucleus (B1950)}& Observing & Alternative \nl
       &  Type   &(km s$^{-1}$)& RA ({\it h m s})& Dec ($^{\circ}$ $'$ $''$)& date & name \nl
\tableline 
Mrk~1  &   2     & 4800& 01 13 19.616 $\pm 0.009$ & +32 49 33.12 $\pm 0.11$& 27/09/90 &NGC 449  \nl
Mrk~3  &   2     & 4110& 06 09 48.419 $\pm 0.023$ & +71 03 10.72 $\pm 0.11$& 01/10/90 &UGC 3426  \nl
Mrk~231&   1     &12300& 12 54 05.004 $\pm 0.012$ & +57 08 38.26 $\pm 0.10$& 30/09/90 &UGC 8058  \nl
Mrk~463&   2     &15150& 13 53 39.858 $\pm 0.007$ & +18 36 57.92 $\pm 0.10$& 30/09/90 &UGC 8850  \nl
\enddata
\end{deluxetable}

\clearpage

\begin{deluxetable}{lccrrrrrrrrr}
\tablecaption{Radio properties of Mrk 1, 231 and 463. The error in the
measured flux densities due to calibration uncertainty is estimated to
be $\sim 10\%$. Values for the magnetic field, energy density and
pressure in the radio plasma were estimated using minimum energy
arguments (Miley 1980), and are given in the following units: $B_{eq}$
$10^{-7}$T; $U_{eq}$ $10^{-8}$ J m$^{-3}$; $P_{eq}$ $10^{-9}$ N
m$^{-2}$. We have assumed spectral indices of $\alpha=-0.7$, where
$S\propto \nu^{\alpha}$.
\label{tbl-2}}
\scriptsize 
\tablewidth{0pt}
\startdata

 & \multicolumn{2}{c}{Radio position} & \multicolumn{2}{c}{18-cm flux} & \multicolumn{1}{c}{rms}& \multicolumn{1}{c}{Component}& &\multicolumn{3}{c}{Equipartition} &\multicolumn{1}{c}{Brightness} \nl
       & \multicolumn{2}{c}{(B1950)} & \multicolumn{2}{c}{density (mJy)} & \multicolumn{1}{c}{noise} & \multicolumn{1}{c}{size} & PA & \multicolumn{3}{c}{parameters} & \multicolumn{1}{c}{temperature} \nl
       & RA ({\it h m s}) & Dec ($^{\circ}$ $'$ $''$) & Peak & Total & \multicolumn{1}{c}{(mJy)} & \multicolumn{1}{c}{(arcsec$^{2}$)} &($^{\circ}$) & $B_{eq}$ &\multicolumn{1}{c}{$U_{eq}$} &\multicolumn{1}{c}{$P_{eq}$} & \multicolumn{1}{c}{(K)} \nl \tableline 

Mrk 1  &01 13 19.5800&+32 49 32.500&29.5&34.0&0.14&$ 0.010\times 0.006$&  78&$5.3$ &$26.4$&$87.9$&$2.9\times10^{8}$\nl
       &             &            &    &    &    &                        &    &      &      &      &                 \nl
Mrk 231&12 54 05.0000&+57 08 38.098&74.7&94.3&0.26&$ 0.018\times 0.006$&  20&$4.2$ &$16.5$&$55.1$&$4.5\times10^{8}$\nl
       &12 54 04.9969&+57 08 38.038&13.2&20.6&0.26&$ 0.023\times 0.012$& 168&$1.9$ &$ 3.5$&$11.6$&$4.2\times10^{7}$\nl
       &12 54 04.9963&+57 08 37.970& 2.2& 2.6&0.26&$\sim0.030\times 0.020$&$\sim$90&$0.8$ &$ 0.5$&$ 1.8$&$2.3\times10^{6}$\nl
       &             &            &    &    &    &                        &    &      &      &      &                 \nl
Mrk 463&13 53 39.8500&+18 36 57.599&34.3&62.7&0.19&$ 0.024\times 0.017$& 148&$2.1$ &$ 4.2$&$13.9$&$8.2\times10^{7}$\nl
       &13 53 39.8488&+18 36 57.546& 6.3&11.6&0.19&$ 0.021\times 0.020$& 159&$1.3$ &$ 1.5$&$ 5.2$&$1.5\times10^{7}$\nl
       &13 53 39.8486&+18 36 57.517& 4.3& 4.4&0.19&$<0.017\times 0.017$& 180&$>1.2$ &$>1.3$&$>4.3$&$>8.5\times10^{6}$\nl
       &13 53 39.8525&+18 36 57.336& 3.6&13.0&0.19&$\sim0.100\times 0.050$&$\sim$180&$0.5$&$ 0.2$&$ 0.7$&$1.4\times10^{6}$\nl
       &13 53 39.8534&+18 36 56.356& 2.3& 7.3&0.19&$\sim0.100\times 0.050$&$\sim$180&$0.4$&$ 0.1$&$ 0.5$&$8.0\times10^{5}$\nl

\enddata

\end{deluxetable}

\clearpage

\begin{deluxetable}{lccrrcrrrrrr}
\tablecaption{Radio properties of Mrk 3 with radio components numbered
as in Figure~5. The beam has a FWHM of $0.027''\times 0.023''$ in PA
$51^{\circ}$ and the RMS noise in the map is
0.08~mJy~beam$^{-1}$. Spectral indices were calculated by comparing
the current data with the 6-cm map of Kukula et al. (1993) (see
Figure~3). Values for the magnetic field, energy density and pressure
in the radio plasma were estimated using minimum energy arguments
(Miley 1980) and are given in the following units: $B_{eq}$
$10^{-7}$T; $U_{eq}$ $10^{-8}$ J m$^{-3}$; $P_{eq}$ $10^{-9}$ N
m$^{-2}$.
\label{tbl-3}}
\scriptsize 
\tablewidth{0pt}
\startdata

 & \multicolumn{2}{c}{Radio position} & \multicolumn{2}{c}{18-cm flux} & \multicolumn{1}{c}{Spectral}& \multicolumn{1}{c}{Component}& &\multicolumn{3}{c}{Equipartition} &\multicolumn{1}{c}{Brightness} \nl
       & \multicolumn{2}{c}{(B1950)} & \multicolumn{2}{c}{density (mJy)} & \multicolumn{1}{c}{index} & \multicolumn{1}{c}{size} & PA & \multicolumn{3}{c}{parameters} & \multicolumn{1}{c}{temperature} \nl
       & RA ({\it h m s}) & Dec ($^{\circ}$ $'$ $''$) & Peak & Total & \multicolumn{1}{c}{$\alpha$} & \multicolumn{1}{c}{(arcsec$^{2}$)} &($^{\circ}$) & $B_{eq}$ &\multicolumn{1}{c}{$U_{eq}$} &\multicolumn{1}{c}{$P_{eq}$} & \multicolumn{1}{c}{(K)} \nl \tableline 
1a & 06 09 48.6105 &+71 03 10.488&  1.8 &  6.7 & ?   &$ 0.100\times0.070$&$\sim180$&$ 0.5$ & $ 0.2    $ & $ 0.7    $ &  $ 5.2\times10^{5}$\nl
1  & 06 09 48.5497 &+71 03 10.608&  2.5 & 45.4 & -2.0&$ 0.270\times0.150$&$\sim90$ &$ 1.4$ & $ 1.8    $ & $ 5.9    $ &  $ 6.1\times10^{5}$\nl
2  & 06 09 48.4988 &+71 03 10.576&  9.7 & 41.8 & -1.0&$ 0.061\times0.036$& 101     &$ 1.6$ & $ 2.5    $ & $ 8.2    $ &  $ 1.0\times10^{7}$\nl
3  & 06 09 48.4742 &+71 03 10.544&  8.7 & 40.3 & -0.5&$ 0.067\times0.036$&  94     &$ 1.1$ & $ 1.2    $ & $ 4.0    $ &  $ 9.0\times10^{6}$\nl
4  & 06 09 48.4249 &+71 03 10.488&  5.4 &  9.4 & -0.1&$<0.038\times0.038$& 180     &$>1.0$ & $>1.0    $ & $>3.3    $ &  $>3.5\times10^{6}$\nl
5a & 06 09 48.3937 &+71 03 10.448&  2.5 &  3.4 & -1.8&$ 0.015\times0.008$& 140     &$ 6.5$ & $39.0    $ & $130.0   $ &  $ 1.5\times10^{7}$\nl
5b & 06 09 48.3838 &+71 03 10.464&  4.7 &  7.5 & -1.8&$ 0.037\times0.017$& 140     &$ 4.0$ & $14.0    $ & $48.0    $ &  $ 6.4\times10^{6}$\nl
5c & 06 09 48.3641 &+71 03 10.456&  2.2 &  4.2 & -1.8&$ 0.023\times0.019$& 149     &$ 4.0$ & $15.0    $ & $49.0    $ &  $ 5.2\times10^{6}$\nl
5d & 06 09 48.3510 &+71 03 10.448&  2.3 &  3.3 & -1.8&$ 0.040\times0.020$&$\sim180$&$ 2.8$ & $7.4     $ & $24.7    $ &  $ 2.2\times10^{6}$\nl
6a & 06 09 48.3214 &+71 03 10.400&  2.1 & 10.7 & -0.9&$ 0.070\times0.040$&$\sim135$&$ 0.9$ & $ 0.7    $ & $ 2.3    $ &  $ 2.1\times10^{6}$\nl
6b & 06 09 48.3116 &+71 03 10.400&  4.1 &  7.5 & -0.9&$ 0.037\times0.034$& 140     &$ 1.1$ & $ 1.2    $ & $ 3.9    $ &  $ 3.2\times10^{6}$\nl
6c & 06 09 48.3050 &+71 03 10.376&  3.2 &  3.9 & -0.9&$ 0.010\times0.007$& 151     &$ 3.2$ & $ 9.4    $ & $31.0    $ &  $ 3.0\times10^{7}$\nl
7  & 06 09 48.2738 &+71 03 10.424&  2.6 &  6.3 & -0.4&$ 0.039\times0.027$&  17     &$ 9.6$ & $ 0.9    $ & $ 2.8    $ &  $ 3.2\times10^{6}$\nl
8a & 06 09 48.1966 &+71 03 10.464& 11.6  & 41.7 & -1.2&$ 0.070\times0.050$&$\sim135$&$ 1.7$ & $ 2.5    $ & $ 8.5    $ &  $ 6.4\times10^{6}$\nl
8b & 06 09 48.1834 &+71 03 10.416& 19.7  & 27.9 & -1.2&$ 0.058\times0.048$&  19     &$ 1.6$ & $ 2.5    $ & $ 8.3    $ &  $ 5.4\times10^{6}$\nl
8c & 06 09 48.1801 &+71 03 10.408& 22.9  & 32.4 & -1.23&$ 0.062\times0.043$&  66     &$ 1.7$ & $ 2.8    $ & $ 9.3    $ &  $ 6.6\times10^{6}$\nl
8d & 06 09 48.1686 &+71 03 10.432& 34.4  &124.3 & -1.2&$ 0.091\times0.044$&  25     &$ 2.1$ & $ 4.1    $ & $14.0    $ &  $ 1.7\times10^{7}$\nl

\enddata
\end{deluxetable}

\clearpage

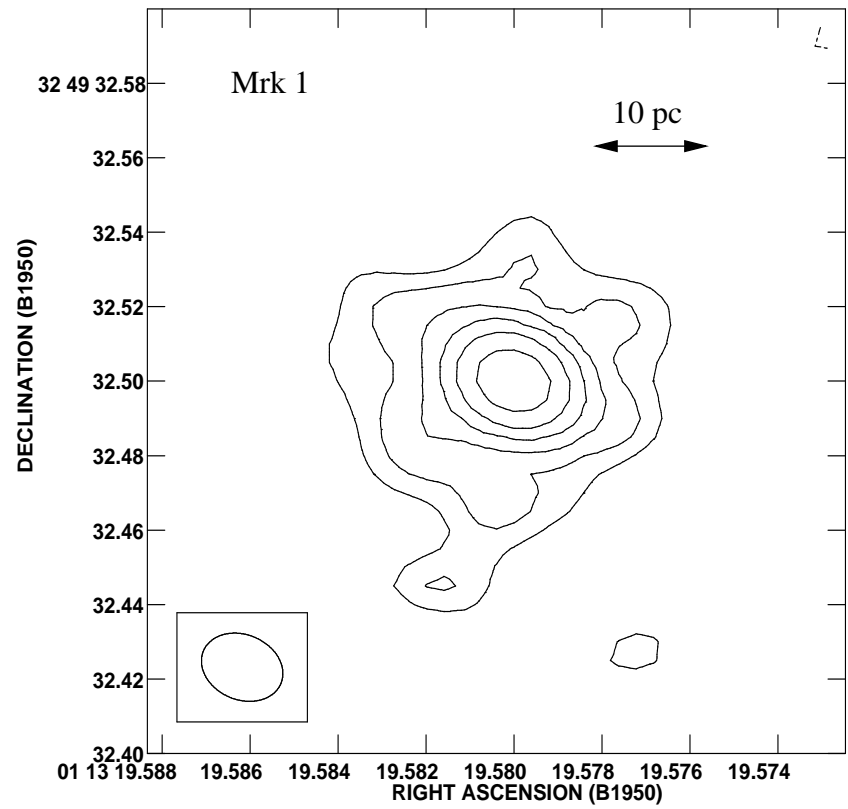
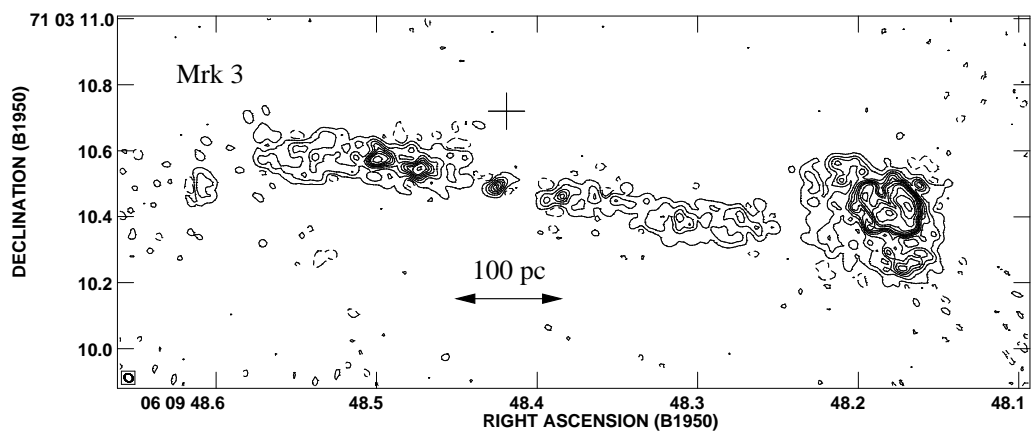
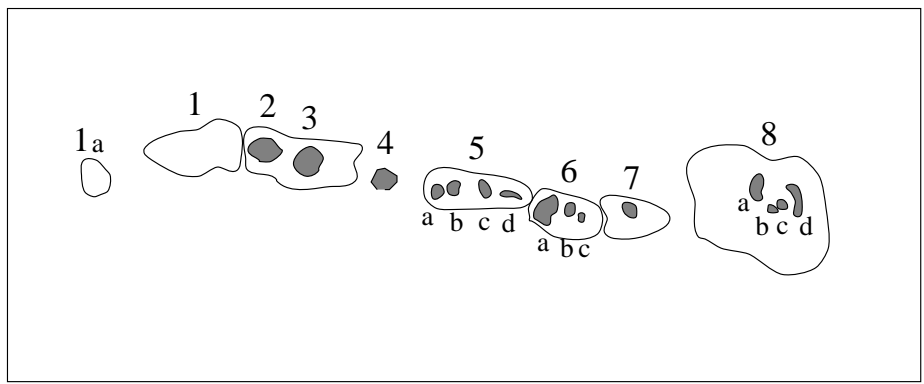
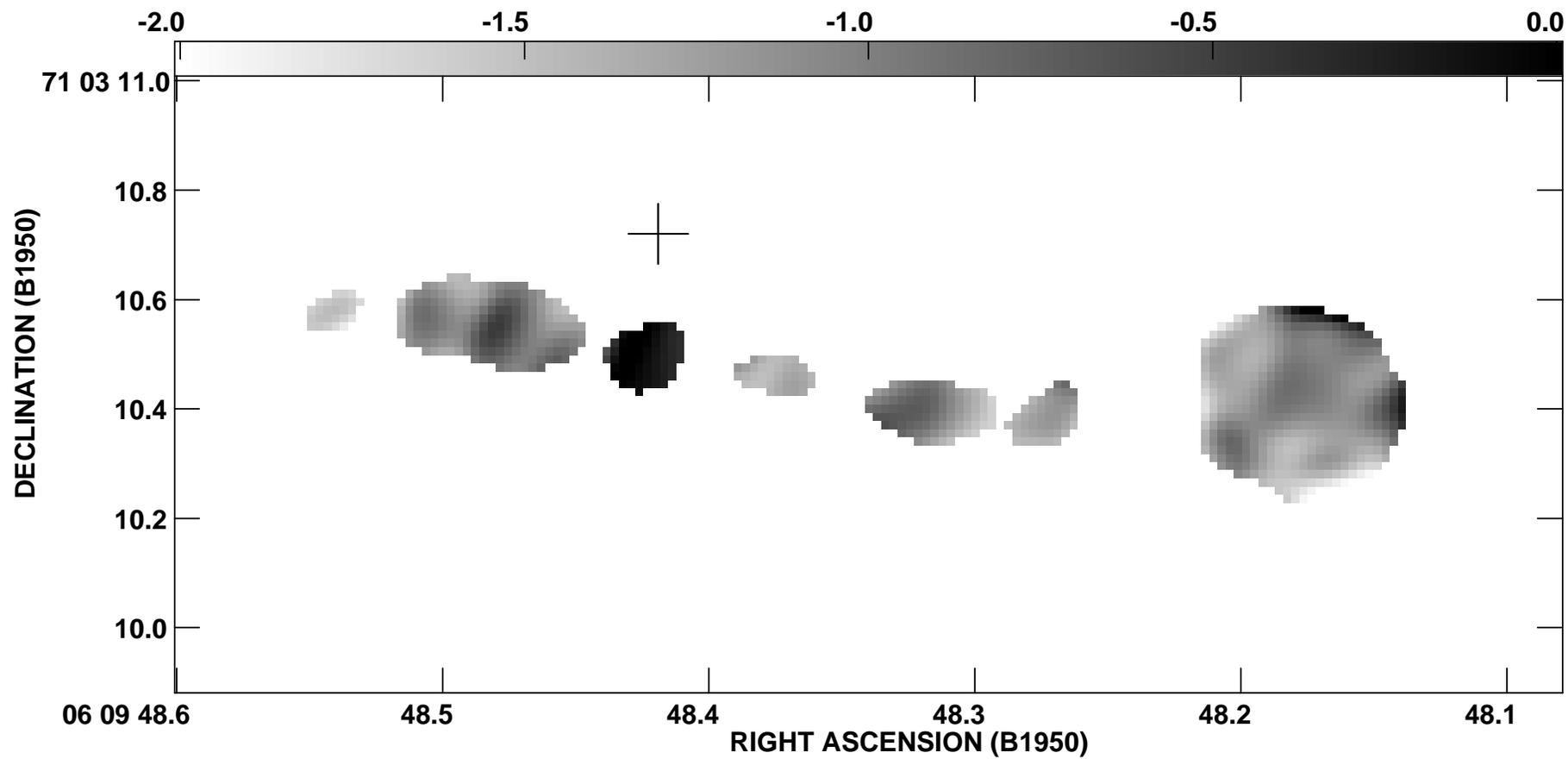
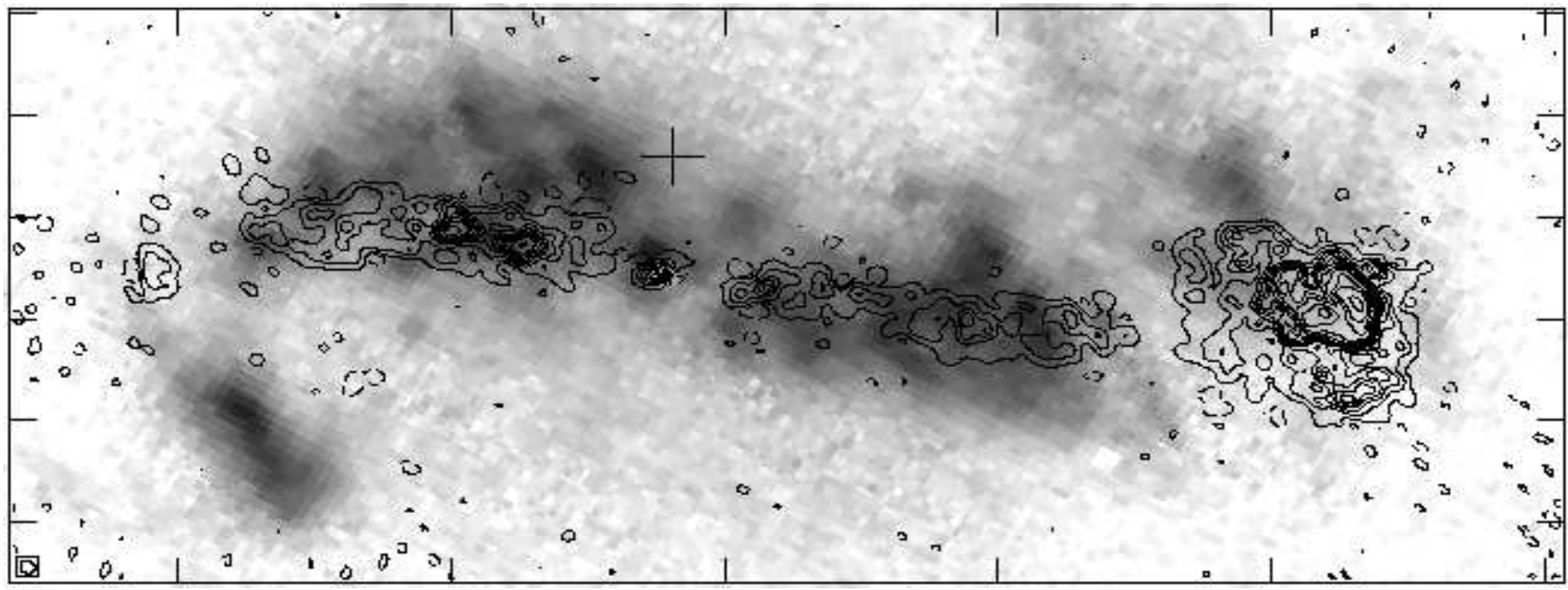
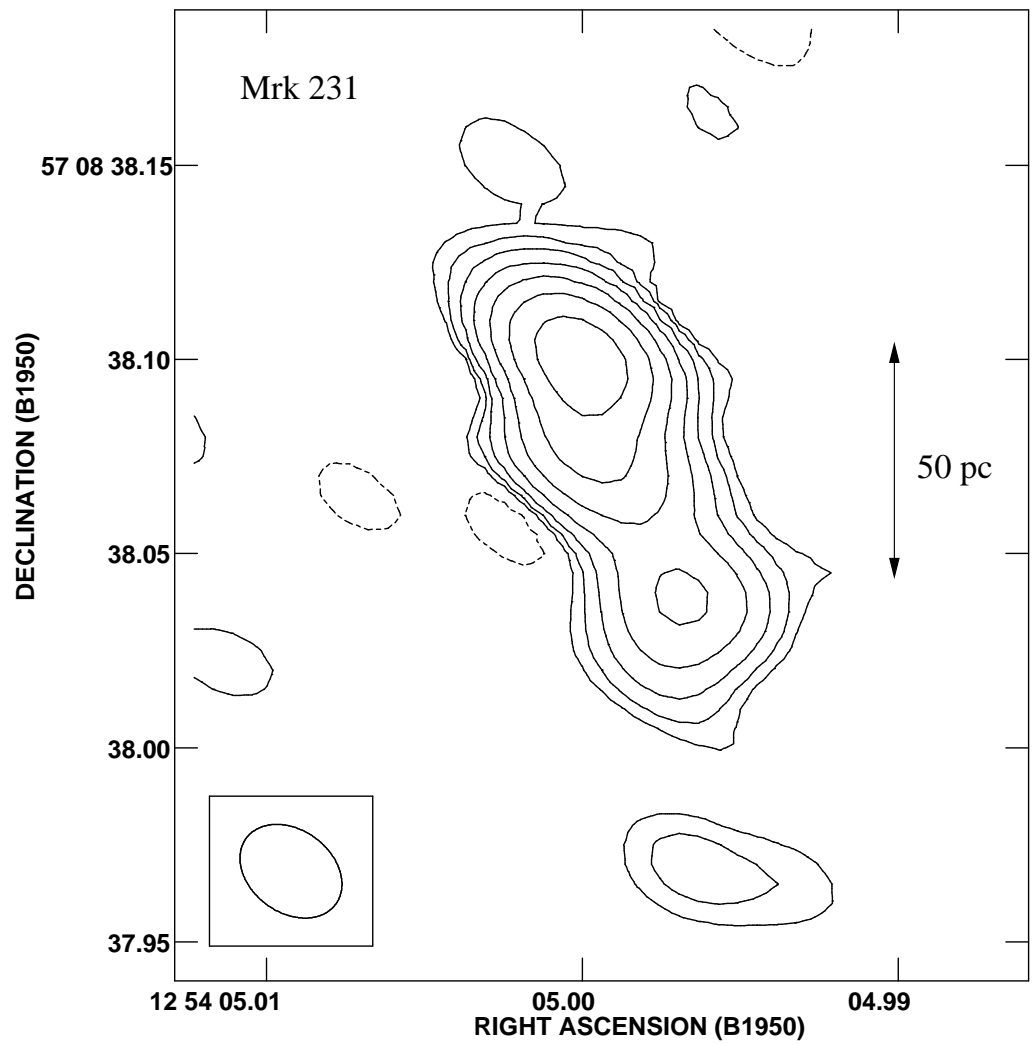
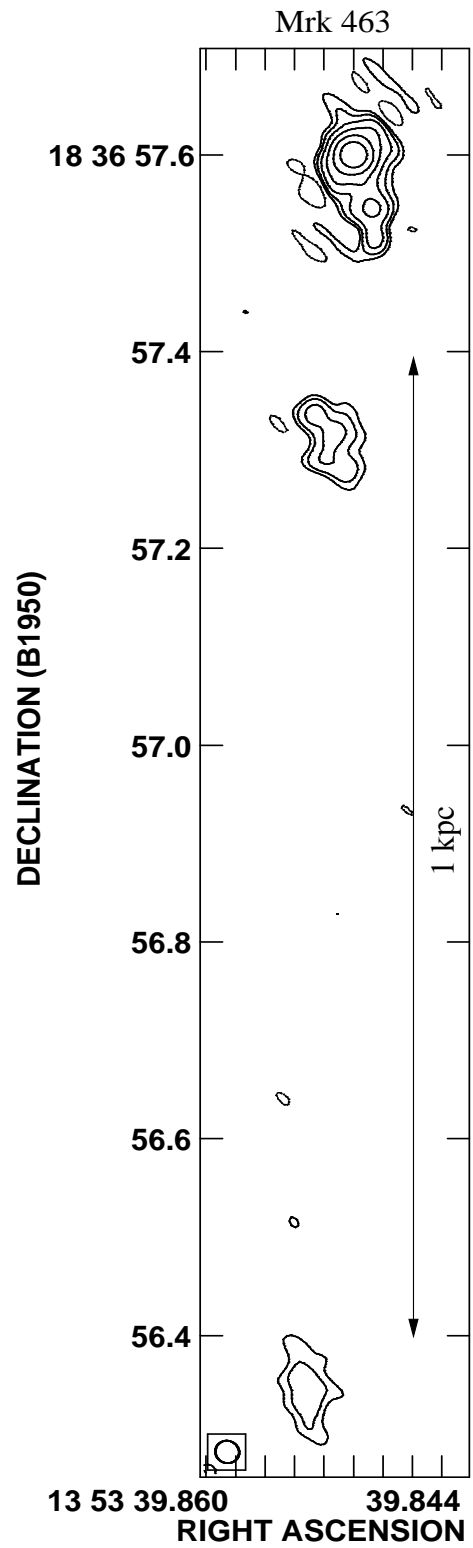
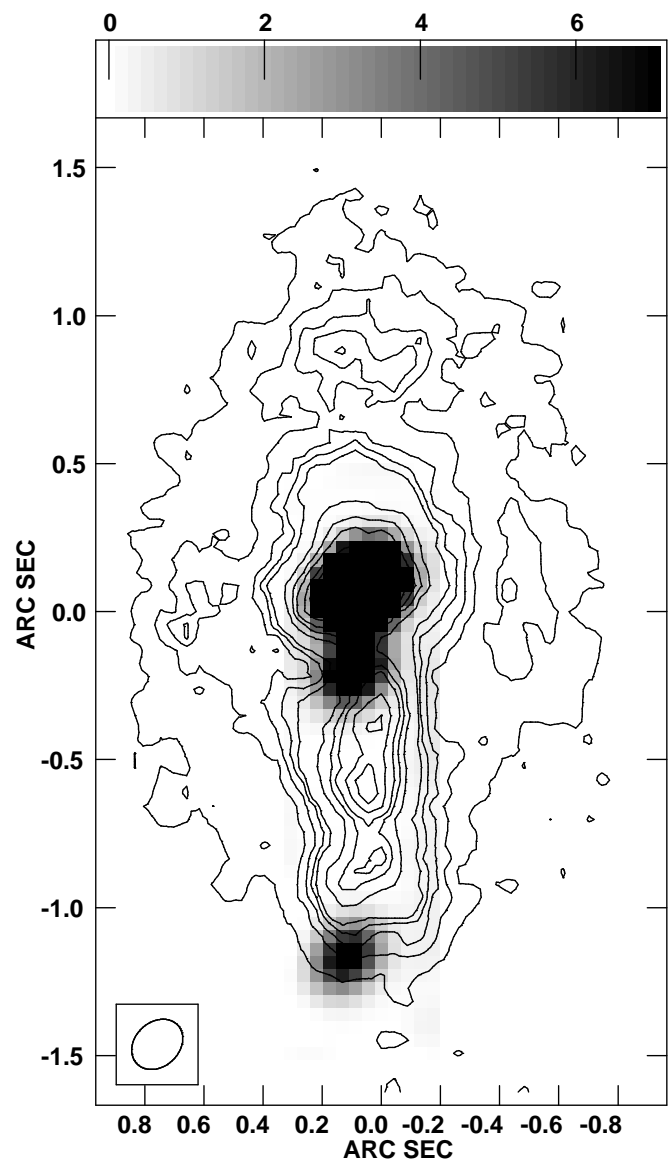
\end{document}